\documentclass[twocolumns]{aa}
\usepackage{graphicx}
\usepackage{txfonts}
\begin{document}
   \title{Early $BVR_{\rm c}I_{\rm c}$ imaging and the discovery of the optical afterglow of GRB 041218}

   \subtitle{}

   \author{Ken'ichi Torii,
          \inst{1}
          Yuuichi Fukazawa,
          \inst{1}  
	  \and	
          Hiroshi Tsunemi
          \inst{1}
          }

   \offprints{Ken'ichi Torii}

   \institute{Department of Earth and Space Science, Graduate School of Science, Osaka University,\\ 1-1 Machikaneyama-cho, Toyonaka, Osaka 560-0043, Japan\\
              \email{torii@ess.sci.osaka-u.ac.jp}
             }

   \date{Received 16 March 2005; accepted *, 2005}

   \abstract{

We report early $BVR_{\rm c}I_{\rm c}$ imaging of the INTEGRAL GRB
041218. The observation started 129 s after the burst and a sequence
of exposures in $V$, $R_{\rm c}$, $I_{\rm c}$, and $B$ bands were
repeated. The optical afterglow is detected in $R_{\rm c}$ and $I_{\rm
c}$ bands while it was not detected in $B$ and $V$.  We find that the
early afterglow deviates from a single power law decay and identify
two characteristic timescales (breaks). Such breaks are theoretically
expected from synchrotron cooling in the relativistic fireball (Sari
et al. 1998) and we crudely constrain the physical condition of the
relativistic shock.

   \keywords{Gamma rays: bursts}
   }
\titlerunning{GRB 041218}
\authorrunning{Torii, Fukazawa, \& Tsunemi}
   \maketitle
%

\section{Introduction}

 The discovery of afterglow emission of gamma-ray bursts (van
 Paradijs et al. \cite{vanparadijs}; Costa et al. \cite{costa})
 opened new diagnostics for quantitative characterization of
 relativistic shock in GRBs. Follow-up observation within minutes of
 the gamma-ray emission is particularly important, since the afterglow
 has information close ($r\leq 0.01$\, pc) to the progenitor.  Early
 optical afterglow has been detected only in a small number of
 occasions, including GRB 990123 (Akerlof et al. \cite{akerlof}), GRB
 021004 (Torii et al. \cite{torii02}; Fox et al. \cite{fox}), GRB 021211
 (Wozniak et al. \cite{wozniak}), and GRB 030418 (Rykoff et
 al. \cite{rykoff}).  In spite of the limited pieces of data, early
 afterglows have shown rich variety of behaviors which provided us
 with crucial information on the physical condition in the vicinity of
 the progenitor. These data brought us with the interesting pictures
 such as forward and reverse shock components (GRB 990123 and
 GRB 021211), continued energy injection from the central engine (GRB
 021004), and large extinction near the massive progenitor (GRB
 030418).

Since all these early detections were made by unfiltered or single
filter CCDs, no information on the afterglow color could be derived in
the previous studies.  In this Letter, we report early $BVR_{\rm c}I_{\rm c}$
imaging and the discovery of the optical afterglow of GRB 041218 with
the Automated Response Telescope (ART).


\section{Observations and Results}

A long (60 s) duration gamma-ray burst, GRB 041218 was detected with
the IBIS/ISGRI instruments onboard the INTEGRAL satellite at 2004
December 18, 15:45:25 UT (Mereghetti et al. 2004) and the burst
position was distributed through IBAS (Mereghetti et
al. \cite{mereghetti03}).  After the earliest successful automated
observation of GRB 030329 (Torii et al. \cite{torii03}; Vanderspek et al. \cite{vanderspek}), we upgraded the ART by using a larger telescope and installing standard photometric filters.
GRB 041218 observations were performed with a 14 inch f/7
Schmidt Cassegrain telescope, equipped with Cousin's $BVR_{\rm c}I_{\rm c}$ filters
(Bessel 1990) in conventional filter wheel and cooled CCD camera
(KAF-1001E).
The first 60 s exposure in $V$ band was started at
15:47:34 UT, or 129 s after the burst. Then a sequence of 60 s
exposures in $R_{\rm c}$, $I_{\rm c}$, and $B$ bands was repeated. When the observation
started, a large fraction of the sky was covered by clouds while the
transparency gradually improved until 20 minutes post-burst. 
After that, the field of view was intermittently covered by
clouds.

\begin{figure}
   \centering
   \includegraphics[angle=0,width=4.25cm]{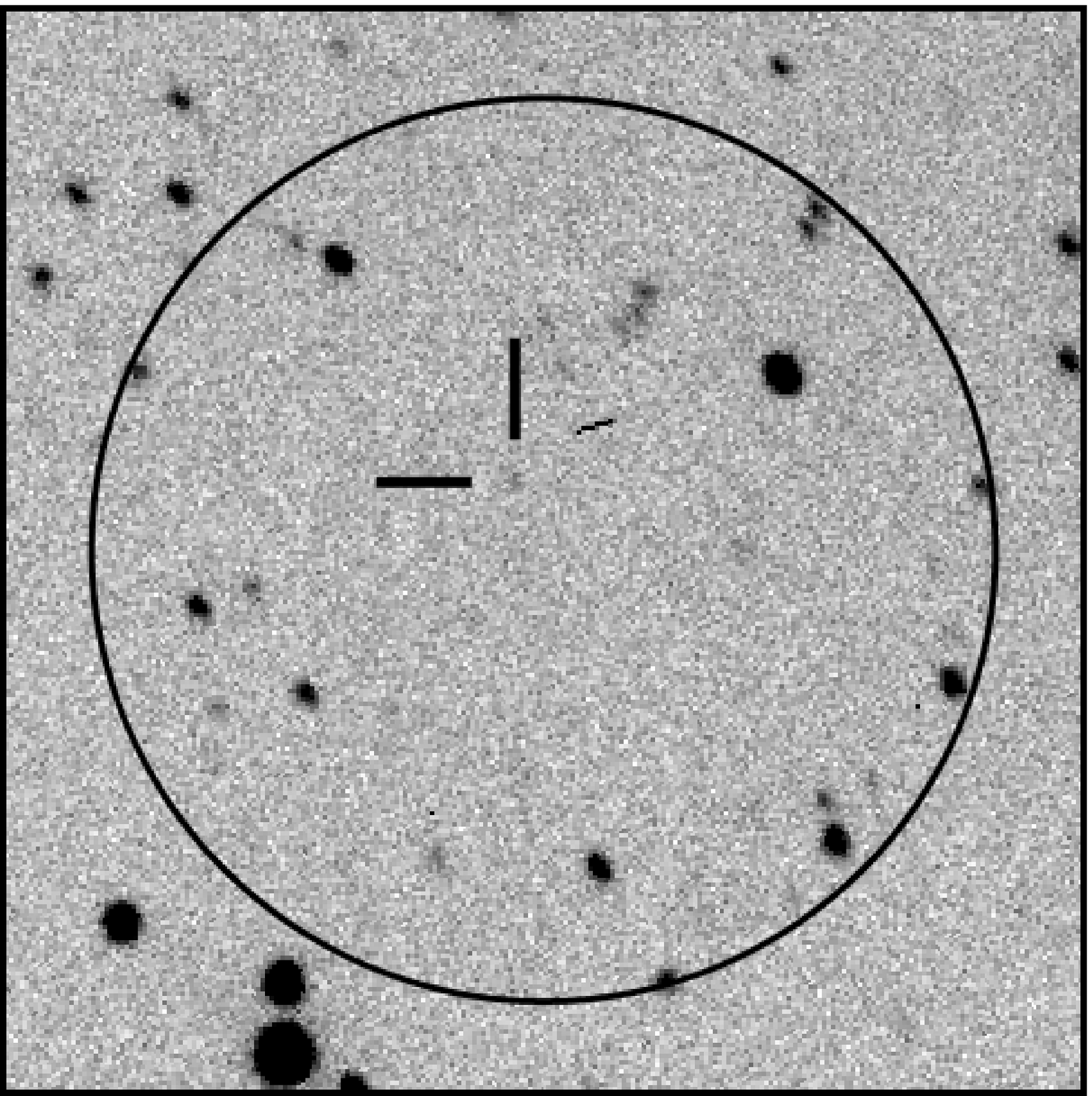}
   \includegraphics[angle=0,width=4.25cm]{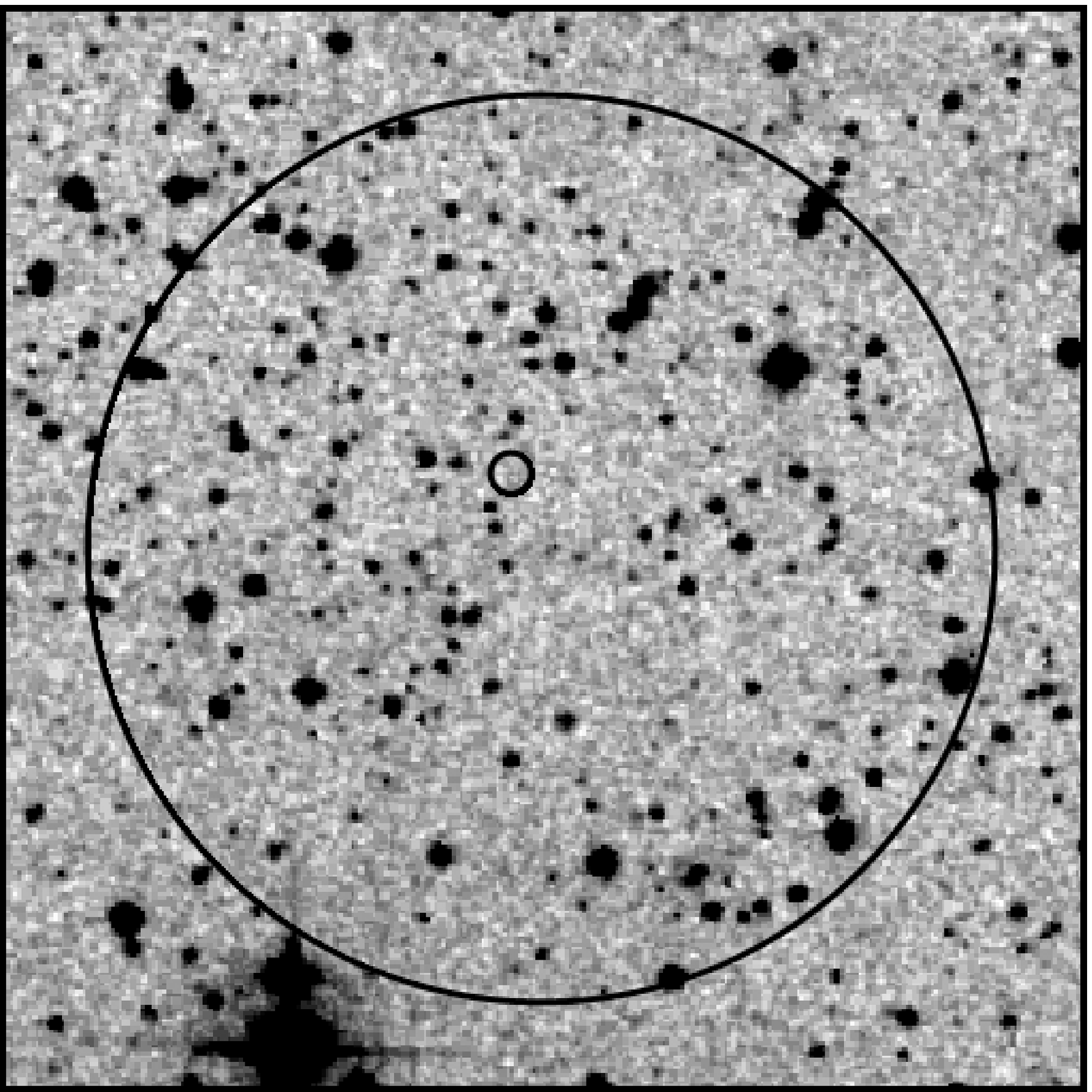}
      \caption{
Left: $6\arcmin\times 6\arcmin$ field around the optical afterglow (at the cross point of vertical and horizontal bars) as imaged by ART in $I_{\rm c}$ band (2$\times$60 s). The circle shows the $2\farcm5$ radius INTEGRAL error region (Mereghetti et al. \cite{mereghetti04}). North is up and East is to the left. The narrow elongated object in the middle of the image is a cosmic ray event. Right: The GRB 041218 field in the digitized Palomar Observatory Sky Survey (DSS2 Red) is shown for comparison. A small circle shows the afterglow position. 
              }
         \label{Fig_image}
   \end{figure}

We stacked four (two $R_{\rm c}$ and two $I_{\rm c}$) frames obtained
on December 18 between 15:57:59 and 16:09:24 UT and visually compared the resulting image with the
digitized Palomar Observatory Sky Survey frame. In this stacked image, we noted a new object with
S/N = 3.8 near the center of the $2\farcm5$ radius error circle (Mereghetti et al. \cite{mereghetti03}) and reported it to
the GCN Circular (Torii \cite{torii04}). An independent identification of this new object and its fading was  reported by Gorosabel et al. (\cite{gorosabel}). 
Figure \ref{Fig_image} shows the GRB 041218 field as imaged by ART and the digitized Palomar Observatory Sky Survey 2 Red for comparison.

\begin{figure}
   \centering
   \includegraphics[angle=-90,width=8.5cm]{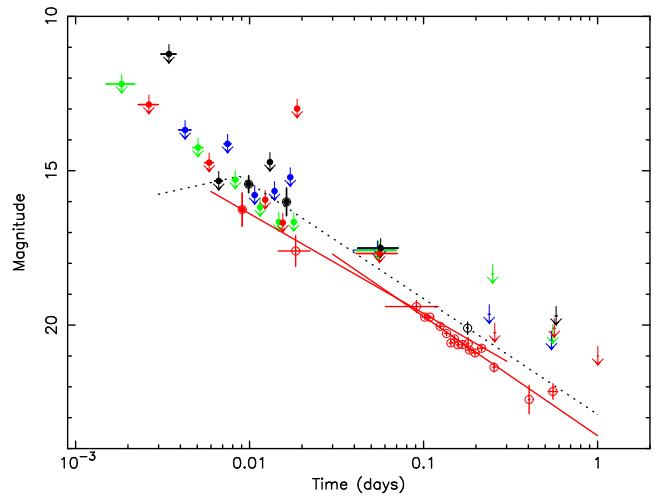}
      \caption{
$BVR_{\rm c}I_{\rm c}$ afterglow light curve of GRB 041218. The horizontal axis shows time from GRB onset. Blue, green, red, and black marks show, $B$, $V$, $R_{\rm c}$, and $I_{\rm c}$ measurements, respectively. In addition to our measurements (filled circles; arrows mark $2\sigma$ upper limits), those reported in GCN circulars (Yonetoku et al. \cite{yonetoku}; Klotz et al. \cite{klotz}; Ferrero et al. \cite{ferrero}; Monfardini et al. \cite{monfardini}; D'Avanzo et al. \cite{davanzo}; Halpern et al. \cite{halpern}; Cenko et al. \cite{cenko}; Mirsa \& Pandey \cite{mirsa}) are plotted with open circles (detections) or arrows (upper limits). See discussion in text for the black and red lines.
              }
         \label{Fig_lc}
   \end{figure}

After dark subtraction and flat fielding, we applied robust source
detection and PSF photometry procedures to the data. 
When the source counts exceeded the $2\sigma$ level we considered it as a detection. In the other cases such $2\sigma$ upper limit was derived.
The afterglow was detected in one $R_{\rm c}$ frame
at $2.2\sigma$ and two $I_{\rm c}$ frames at 4.0 and $2.5\sigma$ level. The derived
magnitudes, calibrated by using the field photometry of Henden
(\cite{henden}), are summarized in Table \ref{table:1}. 
Figure \ref{Fig_lc} shows the $BVR_{\rm c}I_{\rm c}$ light curves combining our data with those reported in the GCN Circulars.

\begin{table}
\caption{Photometric results.}             
\label{table:1}      
\centering                          
\begin{tabular}{c c c c l}        
\hline\hline                 
Start Time & Mean Epoch & Magnitude* & Filter & Exposure \\
(UT)       & (days)     &           &        & (s) \\ \hline 
15:47:34 & 0.00184 &  $>$12.19&$V$ & 60\\ 
15:48:43 & 0.00263 &  $>$12.85&$R_{\rm c}$ & 60\\
15:49:52 & 0.00343 & $>$11.22&$I_{\rm c}$ & 60\\

15:51:03    & 0.00425 &    $>$13.68&$B$ & 60\\ 
15:52:12    & 0.00505 &    $>$14.25&$V$ &60\\
15:53:21    & 0.00585 &    $>$14.73&$R_{\rm c}$ &60\\
15:54:29    & 0.00664 &    $>$15.33&$I_{\rm c}$ &60\\

15:55:41    & 0.00747 &    $>$14.13&$B$ &60\\ 
15:56:50    & 0.00827 &    $>$15.29&$V$ &60\\
15:57:59    & 0.00907 &    16.25$\pm$0.54&$R_{\rm c}$ &60\\
15:59:08    & 0.00987 &    15.43$\pm$0.27&$I_{\rm c}$ &60\\

16:00:19    & 0.01069 &    $>$15.79&$B$ &60\\ 
16:01:28    & 0.01149 &    $>$16.18&$V$ &60\\
16:02:37    & 0.01229 &    $>$15.94&$R_{\rm c}$ &60\\
16:03:46    & 0.01309 &    $>$14.72&$I_{\rm c}$ &60\\

16:04:57    & 0.01391 &    $>$15.66&$B$ &60\\ 
16:06:06    & 0.01471 &    $>$16.66&$V$ &60\\
16:07:15    & 0.01550 &    $>$16.68&$R_{\rm c}$ &60\\
16:08:24    & 0.01630 &    16.00$\pm$0.45&$I_{\rm c}$ &60\\

16:09:35    & 0.01712 &    $>$15.21&$B$ &60\\ 
16:10:46    & 0.01795 &    $>$16.66&$V$ &60\\
16:11:56    & 0.01876 & $>$12.99 & $R_{\rm c}$ &60\\
16:42:10 & 0.05425 &$>$17.62&$B$&60\,$\times$\,9 \\ 
16:43:19 & 0.05505 &$>$17.58&$V$&60\,$\times$\,10 \\ 
16:44:28 & 0.05585 &$>$17.50&$R_{\rm c}$&60\,$\times$\,10 \\ 
16:45:37 & 0.05664 &$>$17.68&$I_{\rm c}$&60\,$\times$\,10 \\ 
\hline                                   
\end{tabular}
\\ $*$ Upper limits are at $2\sigma$ level.
\end{table}

\section{Discussion}

The afterglow light curve of GRB 041218 shows, roughly, a power law
decay. Here, we try to model the light curve as synchrotron radiation from 
the relativistic fireball.

Fitting the $R_{\rm c}$ band light curve with the function ${\rm Mag} = -2.5 \cdot \log t^a + b$ ($t$ in days) gives
$a_0=-1.49\pm0.10$ and $b_0=23.47\pm 0.23$ with $\chi^2/{\rm dof} =
32.7/16$. The errors for $a$ and $b$ are 90\% confidence values. A large $\chi^2$ value is mainly due to fluctuations at
$0.1\stackrel{<}{_\sim} t \stackrel{<}{_\sim} 0.3$ days while this function
marginally overestimate the flux in the first two $R_{\rm c}$ detections,
suggesting that the flux decay rate steepened at around $0.04-0.08$
days. If we fit the $R_{\rm c}$ data at $t>0.05$ days, a steeper index
$a_2=-1.54\pm0.12$ and $b_2=23.58\pm0.26$ are obtained with
$\chi^2/{\rm dof} = 30.2/14$. 

For the $I_{\rm c}$ band, only three detections are available and they are
well described with a single power law with $a_{3}=-1.49\pm0.16$ and
$b_{3}=22.89\pm0.58$.  This function is drawn as dotted black line in Figure
\ref{Fig_lc} for $t \ge 0.00987$ days.  The decay index in $I_{\rm c}$ band is fully
consistent with that for the $R_{\rm c}$ band.  However, the upper
limit at 0.00664 days is significantly (0.59 mag) below the extrapolation
from this power law.  This means that the flux decay with $a_3=-1.49$
started just before the earliest detection at 0.00987 days.  By using the upper limit at 0.00664
days, the decay index must have been $a > -0.18$ at $t<0.00987$ days
and suddenly changed to $a_3=-1.49$ within $\sim 5$ minutes.

We note that the observed changes in the decay indices (at $0.00664<t<0.00987$ days in $I_{\rm c}$ and at $0.04<t<0.08$ days in $R_{\rm
c}$) are theoretically expected for synchrotron radiation from 
relativistic fireball (e.g., Sari et al. 1998). According to this model, the early optical afterglow is
expected to show breaks. The first one, $t_m$, corresponds to the
transit in the optical band of the typical electrons synchrotron
frequency for minimum Lorentz factor, $\gamma_m$. The second one,
$t_c$, is the break corresponding to the transit of the cooling
frequency in the observing band. Therefore, we interpret
$t_m \simeq 0.00664-0.00987$ days and $t_c \simeq
0.04-0.08$ days.

 We may constrain the power law index of electron 
number distribution, $p$, from the $R_{\rm c}$ band light curve at $t>0.05$
days. From the relation $a = (2-3p)/4$ and $a_2=-1.54\pm0.12$, we derive
$p=2.72\pm0.16$.  Before the cooling break at $t_c$, the decay index
is expected to be $a = 3(1-p)/4$. Therefore, we draw a power law decay
with $a_1=-1.29$ that starts with our first $R_{\rm c}$ detection at
$t=0.00907$ days.  These two power law functions are shown
in Figure \ref{Fig_lc} by red lines; they intersect at $t_c \simeq 
7.0\times 10^{-2}$ days.

We then constrain $t_m$ so that the limit in $I_{\rm c}$ at $t=0.00664$ days is
consistent with the model function. In the early part of the afterglow
light curve, the flux is expected to increase with a power law index
$a=+1/2$ (Sari et al. \cite{sari}). This constraint leads to a rising light
curve at $t\stackrel{<}{_\sim} 8.7\times 10^{-3}$ days, as drawn in
Figure \ref{Fig_image}.  
The existence of $t_m$ was originally invoked to conserve
the total energy above $\gamma_m$ and it is
unclear if such a clear-cut lower boundary exists in the accelerated electrons
distribution.  The fact that the early afterglow seems to have
brightened in a short timescale ($\sim$minutes)  suggests
 that there is a distinct lower limit in the electron
energy distribution at $\gamma_m$.

For the other parameters, we can not solve the degeneracy due to the
limited pieces of information. Instead, we try to build some
trial parameter sets in the framework of the Sari et al. (\cite{sari}) model with adiabatic evolution. An arbitrary assumption of isotropic burst
kinetic energy $E=1.0\times 10^{52}$~ergs and the fraction of energy
density in the magnetic field 
$\epsilon_B=3.0\times 10^{-2}$ 
gives the astrophysically
reasonable set of other parameters, the fractional energy density of
accelerated electrons, 
$\epsilon_e = 0.053$, 
and the density of the
surrounding medium, 
$n=4.2\, {\rm cm^{-3}}$.

The above discussion applies for constant density around the
progenitor. The environment of long duration gamma-ray bursts are
likely surrounded by stellar wind from a massive progenitor.  In the
framework of the Chevalier \& Li (2000) model, the early flux rise in
GRB 041218 afterglow may be explained as due to the
clearance of the synchrotron self absorption.  In this case (Case B in
Chevalier \& Li (2000)), the early afterglow brightens as $\propto
t^{7/4}$. But again, the model parameters can not be fully constrained and we
can not distinguish whether the interstellar or the circumstellar
model gives the better description.

By interpolating $R_{\rm c}$ and $I_{\rm c}$ measurements, we may
derive the afterglow color as $(R_{\rm c}-I_{\rm c})_1 \simeq
1.0\pm0.6$ at $t=0.00907$ days and $(R_{\rm c}-I_{\rm c})_2 \simeq
0.6\pm0.3$ at $t=0.1791$ days. These values are within the range
($R-I=0.46\pm 0.18$, ${\rm \check{S}}$imon et al. \cite{simon})
measured in other afterglows.  Although the Galactic extinction in the
burst direction is estimated as E($B-V$)=0.61 or A($R_{\rm
c}$)$-$A($I_{\rm c}$)=0.45 from dust emission (Schlegel et
al. \cite{schlegel}), the optical star count suggests a lower value
(A($V$)$<$1.0 or A($R_{\rm c}$)-A($I_{\rm c}$) $<$ 0.22) (Dobashi et
al. \cite{dobashi}, figure 12).  The fact that the GRB 041218
afterglow has an $R_{\rm c}-I_{\rm c}$ color similar to that of other unextincted
afterglows suggests that the lower extinction is likely correct.
The $V-R_{\rm c} > -0.27$ at $t=0.01471$ days is not that strong
constraint and consistent with those observed in other afterglows
(${\rm \check{S}}$imon et al. \cite{simon}).

\section{Conclusions}

We presented early $BVR_{\rm c}I_{\rm c}$ imaging results and the discovery of the optical afterglow of GRB 041218. The afterglow light curve was found to deviate from a single power law decay and two characteristic timescales, $t_m\simeq 8.7\times10^{-3}$ days and $t_c \simeq 7.0\times
10^{-2}$ days, were identified.
We showed that the standard fireball model gives simple enough account for the observed data.

 In the early phase of GRB
afterglow, a variety of unexplored physical processes, such as interaction of
$e^{\pm}$ pairs with the swept-up materials are expected (Beloborodov
2005). We expect that further $BVR_{\rm c}I_{\rm c}$ follow-up observations with the ART for
the INTEGRAL, HETE-2, and Swift GRBs will give insights into such
phenomena.

\begin{acknowledgements}
This work is partly supported by a Grant-in-Aid for Scientific Research
by the Ministry of Education, Culture, Sports, Science and Technology of
Japan (15740129 and 16002004).  This study is also carried out as part of the 21st
Century COE Program, \lq{\it Towards a new basic science: depth and
synthesis}\rq.  

The Digitized Sky Survey was produced at the Space Telescope Science
Institute based on photographic data obtained using the Oschin Schmidt
Telescope on Palomar Mountain. The Second Palomar Observatory Sky
Survey was made by the California Institute of Technology.

\end{acknowledgements}

\end{document}